\documentclass[notitlepage,superscriptaddress,showpacs,nobalancelastpage,twocolumn,aps,prb]{revtex4-1}
 \usepackage[english]{babel}
\RequirePackage[T1]{fontenc}
\RequirePackage{times} 

\usepackage{siunitx} 
\usepackage[italicdiff]{physics}
\usepackage{amsfonts}
\usepackage{subfigure}
\usepackage{amsmath}
\usepackage{amssymb}
\usepackage{graphicx,epstopdf}
\usepackage{xspace}
\usepackage{array}
\usepackage[hidelinks]{hyperref}
\usepackage{xcolor}
\usepackage{my_symbols}
\usepackage{CJK}
\usepackage{bm}
\usepackage{verbatim}
\usepackage{tikz}
\usepackage[normalem]{ulem}
\usepackage{lipsum}

\begin{document}
\begin{CJK*}{UTF8}{gbsn} 

\title{Geometric magnonics with chiral magnetic domain walls}

\author{Jin Lan (兰金)}
\affiliation{Center for Joint Quantum Studies and Department of Physics, School of Science, Tianjin University, 92 Weijin Road, Tianjin 300072, China}
\affiliation{Department of Physics and State Key Laboratory of Surface Physics, Fudan University, Shanghai 200433, China}

\author{Weichao Yu (余伟超)}
\affiliation{Institute for Materials Research, Tohoku University, Sendai 980-8577, Japan}
\affiliation{Department of Physics and State Key Laboratory of Surface Physics, Fudan University, Shanghai 200433, China}

\author{Jiang Xiao (萧江)}
\email[Corresponding author:~]{xiaojiang@fudan.edu.cn}
\affiliation{Department of Physics and State Key Laboratory of Surface Physics, Fudan University, Shanghai 200433, China}
\affiliation{Institute for Nanoelectronics Devices and Quantum Computing, Fudan University, Shanghai 200433, China}

\begin{abstract}
Spin wave, the collective excitation of magnetic order, is one of the fundamental angular momentum carriers in magnetic systems.
Understanding the spin wave propagation in magnetic textures lies in the heart of developing pure magnetic information processing schemes.
Here we show that the spin wave propagation across a chiral domain wall follows simple geometric trajectories, similar to the geometric optics.
And the geometric behaviors are qualitatively different in normally magnetized film and tangentially magnetized film.
We identify the lateral shift, refraction, and total reflection of spin wave across a ferromagnetic domain wall.
Moreover, these geometric scattering phenomena become polarization-dependent in antiferromagnets, indicating the emergence of spin wave birefringence inside antiferromagnetic domain wall.
\end{abstract}
\maketitle
\end{CJK*}

\section{Introduction}
Spin wave, the propagating disturbance of ordered magnetization, is one of the basic excitations in magnetic systems.
As an alternative spin current carrier besides the spin-polarized electrons\cite{kajiwara_transmission_2010}, the spin wave manipulation is not only important for fundamental physics, but also attractive for industrial applications \cite{kruglyak_magnonics_2010,chumak_magnon_2015}.
Due to recent developments in experimental techniques, including excitation in short wavelength \cite{liu_long_2018} and large amplitude \cite{han_mutual_2019}, propagation in long distance \cite{liu_long_2018,cornelissen_long-distance_2015} as well as detection with high sensibility \cite{lee-wong_Nanoscale_2020,li_spin_2020},
magnonics as a discipline devoted to manipulate spin wave is receiving increasing interests\cite{brataas_spin_2020,yu_magnetic_texture_2020a}.

Multiple approaches have been developed to control the spin wave, such as applying external magnetic field \cite{kittel_Theory_1948}, current \cite{demidov_magnetization_2017} and heat \cite{xiao_theory_2010}, as well as coupling with microwave \cite{harder_cavity_2018} and acoustic wave\cite{weiler_spin_2012a}.
Restricted by the external sources introduced in these approaches, the spin wave devices are typically difficult to miniaturize.
An alternative approach is using magnetic texture widely existing in magnetic materials, including domain wall, magnetic vortex, magnetic Skyrmion etc.
Since both magnetic texture and spin wave are of intrinsic magnetic nature, thus can coexist in single magnetic material, and intimately interact with each other.
Using magnetic texture to store information, and spin wave to process information, pure magnetic computing schemes can be developed \cite{lan_spin-wave_2015,han_mutual_2019,yu_magnetic_2020,yu_magnetic_texture_2020a}.

The influence of magnetic texture on spin wave are mostly focused on the wave aspects of the spin wave, including its amplitude, phase and polarization.
The domain wall naturally act as the waveguide for spin wave \cite{garcia-sanchez_narrow_2015,lan_spin-wave_2015,wagner_magnetic_2016,sluka_emission_2019}, and magnetic vortex functions as spin wave emitter \cite{wintz_magnetic_2016}.
A Mach-Zehnder interferometer for spin wave can be constructed, by preparing domain wall in one arm of a two-arm structure \cite{hertel_domain-wall_2004,buijnsters_dzyaloshinskii-moriya_2015}.
In presence of the Dzyaloshinskii-Moriya interaction (DMI), an antiferromagnetic domain wall naturally serves as spin wave polarizer and retarder \cite{lan_antiferromagnetic_2017-1}.
However, the existing investigations on spin wave trajectory, dictating the particle aspect of the spin wave, rely heavily on the wave-based equations \cite{iwasaki_theory_2014,schutte_magnonskyrmion_2014a} or effects \cite{yu_magnetic_2016,stigloher_Snell_2016,gruszecki_goos-hanchen_2017,wang_GoosH_2019}, with straightforward and quantitative trajectory analysis missing.

In this work, we systematically investigate the spin wave scattering by chiral domain wall in both normally magnetized film and tangentially magnetized film.
Based on the semiclassical analysis and micromagnetic simulations, we identify various geometrical relations between incident and out-going spin wave beams, including lateral shift, refraction and the total reflection, similar to its optical counterpart.
And these geometrical magnonic phenomena become polarization dependent in when extending to antiferromagnetic environment.
The geometrical magnonics as demonstrated in this work, offers us simple yet intuitive paradigms in constructing magnonic devices of different functionalities.

This paper is organized as follows.
In Sec. II, a semiclassical scheme that describes the spin wave scattering by chiral domain wall is established.
Based on the semiclassical trajectory analysis and the micromagnetic simulations, various geometrical magnonic phenomena in normally and tangentially magnetized films are then demonstrated, and further understanding by magnonic Snell's law is provided.
Sec. III is devoted to extension of above the geometric magnonic phenomena to antiferromagnetic environment.
The spin wave constriction by chiral domain wall is presented in Sec. IV, and a short conclusion is drawn in Sec. V.

\begin{figure*}[ht]
\centering
\includegraphics[width=\textwidth]{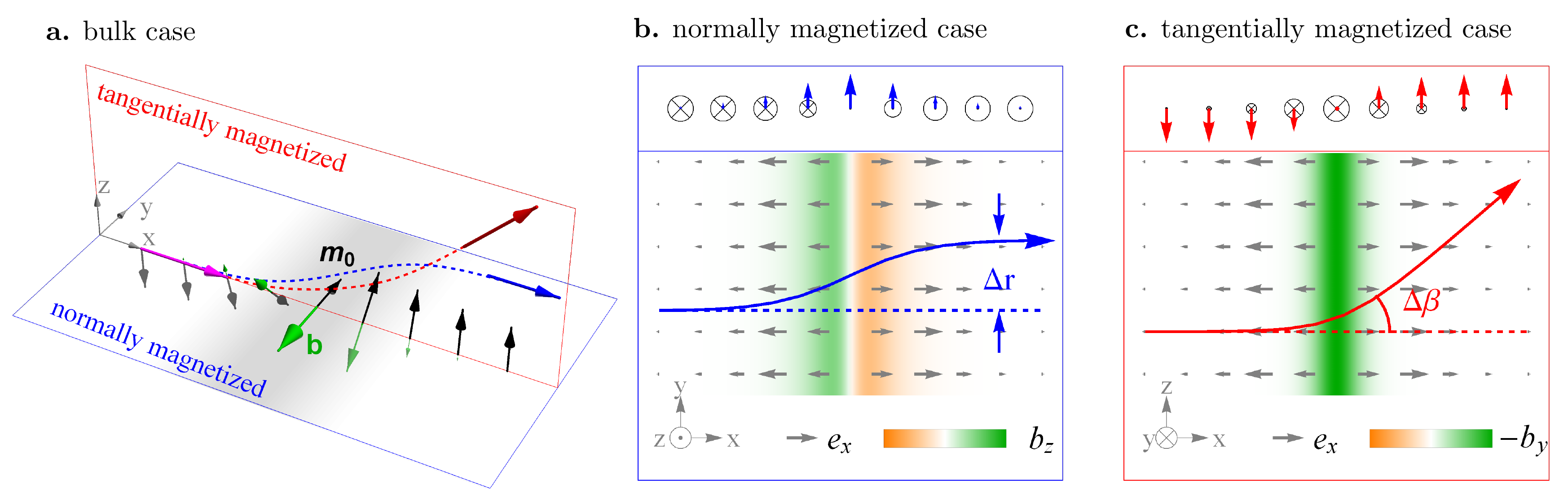}
\caption{ {\bf Schematics of spin wave scattering across a chiral domain wall}.
(a) is in bulk material, and (b)(c) are in normally/tangentially magnetized films, which are slice cuts of (a) in $x-y$ and $x-z$ plane respectively.
(a) A magnetic domain wall along $x$ direction and has translational invariance in the $y$-$z$ plane.  The black/green arrows denote domain wall magnetization $\mb_0$ and fictitious magnetic field $\bb$ respectively, and the gray-scale background is for the scalar potential $\phi$. The blue/red slicing cut of the 3-dimensional magnetic texture corresponds to domain walls in a normally magnetized and tangentially magnetized 2-dimensional magnetic film.
The magenta arrow denotes the incident spin wave beam, and the blue/red arrows denote the out-going beams in $x-y$ and $x-z$ planes respectively.
In (b)(c), the gray arrow depicts the electric field, the green/orange colors encode the positive/negative magnetic field, and the blue/red lines are the typical trajectories for normally incident spin wave on the domain wall.
In the upper region, the magnetization distribution are depicted by arrows, with the in-plane magnetizations highlighted in blue/red color.
}
\label{fig:bend_sch}
\end{figure*}

\section{Basic model}
\subsection{Spin wave dynamics in chiral domain wall}
Consider a ferromagnetic system with its magnetization direction denoted by unit vector $\mb$, then its magnetic dynamics is governed by Landau-Lifshitz-Gilbert (LLG) equation
\begin{equation}
 \label{eqn:LLG}
 \dot{\mb} = -\gamma \mb \times \bh +\alpha \mb \times \dot{\mb} ,
\end{equation}
where $\dot{\mb}\equiv \partial_t \mb$, $\gamma$ is the gyromagnetic ratio, and $\alpha$ is the Gilbert damping constant.
The effective magnetic field $\bh=-\delta u[\mb]/\delta \mb$, where
$u[\mb]= -\half\qty[ K(1- (\mb\cdot \hbe_z)^2) + A(\nabla \mb)^2 +D \mb\cdot (\nabla \times \mb ) ]$
is the magnetic energy density with $K$ the easy-axis anisotropy along $\hbz$, $A$ the exchange coupling constant, and $D$ the Dzyaloshinskii-Moriya interaction (DMI) constant.

The total magnetization naturally divides into the static and dynamical parts: 
$\mb(\br,t)=\mb_0(\br)+\delta \mb(\br,t)$, where $\mb_0(\br)$ represents the static magnetic texture, and $\delta\mb(\br,t)$ is the dynamical spin wave excitation.
In spherical coordinate with $\be_r\equiv\mb_0(\br)$ and the accompanying two transverse directions $\hbe_{\theta,\phi}$, the spin wave is expressed as $\delta\mb(\br,t)=m_\theta(\br,t) \hbe_\theta+m_\phi(\br,t)\hbe_\phi$, or equivalently as a complex field $\psi(\br,t)=m_\theta(\br,t)-i m_\phi(\br,t)$.
We define $u_0\equiv u[\mb_0]$ as the energy density due to the static background $\mb_0$, and $\delta u \equiv u[\mb]-u_0$ as the energy density due to the spin wave excitation.

For a homogenous domain with its static background magnetization $\mb_0(\br) = \pm \hbz$, we have $u_0 = 0$.
A domain wall arises when two different domains meet, and has finite energy $u_0 > 0$.
Without loss of generality, we suppose that the domain wall magnetization varies along $x$-axis, \ie $\mb_0(x)$ rotates continuously from $-\hbz$ to $+\hbz$ along $x$-axis with $\mb_0(\pm\infty) = \pm\hbz$, and is translational invariant along $y/z$-axis.
Due to the DMI, the magnetization inside the domain wall is enforced to rotate counter-clockwisely along the advancing direction $-\hbz\to +\hby\to +\hbz$ along $x$-axis, as shown in Fig. \ref{fig:bend_sch}(a).
Upon this chiral domain wall, the spin wave dynamics is governed by a Schr\"{o}dinger-like equation \cite{lan_spin-wave_2015,lan_skew_2020}
\begin{align}
 \label{eqn:sw_eom_FM}
 i \dot{\psi} = \gamma \qty[ (-i\nabla+\ba)^2 + K -\phi]\psi,
\end{align}
where the vector potential $\ba= \tilde{D}\mb_0$ with $\tilde{D}= D/2A$ \cite{van_hoogdalem_magnetic_2013-1,kim_tunable_2019}, and the scalar potential $\phi= 2u_0$
is caused by the reduction of domain wall energy density $u_0$ by spin wave excitation, which reduces the local magnetization $\mb_0\to \mb_0\sqrt{1-\delta \mb\cdot \delta \mb} $ due to unity condition $|\mb|=1$.

\subsection{Semiclassical description}
To investigate the spin wave scattering behavior by a chiral domain wall, we consider a spin wave packet $\psi[\br(t)]$ centered at position $\br$ in a given time $t$.
In momentum space, we assume that the wave packet is centered at $\bq$, and has sufficient broadening such that the packet is strictly confined in real space.
Following the time-dependent variable principles \cite{sundaram_wave-packet_1999-1, xiao_berry_2010, lan_skew_2020}
in semiclassical approach proposed by Sundaram \emph{et al},
the Lagrangian density corresponding to \Eq{eqn:sw_eom_FM} reads
\begin{align}
\label{eqn:sw_Lag}
\cL= \bk\cdot \dot{\br}-\ba\cdot \dot{\br} -\omega,
\end{align}
where $\omega= \gamma (A\bk^2+K-\phi)$ is the local spin wave frequency with the canonical momentum $\bk=\bq+\ba$. 
Invoking the Euler-Lagrangian rule on \Eq{eqn:sw_Lag}, the dynamics of the spin wave packet is then governed by
\begin{align}
\label{eqn:eom_wp}
m_\ssf{FM}\ddot{\br} = -\be - \dot{\br}\times \bb,
\end{align}
where $m_\ssf{FM}= 1/(\partial_\bk^2 \omega)=1/(2\gamma A)$ is the effective mass for the spin wave in ferromagnets, and $\bv \equiv\dot{\br}=\partial_\bk \omega_0= \bk/m_\ssf{FM}$ is the spin wave velocity.
Here $\be = -\partial_x\phi\hbx$ and $\bb =\nabla \times \ba$ are the fictitious electromagnetic fields induced by inhomogeneous magnetic texture, \ie the chiral domain wall here.
The semiclassical equation \eqref{eqn:eom_wp} is similar to the eikonal equations for the phase of propagation wave, which has been widely used in studying the trajectory of light and gravitational wave \cite{lifshitz_1999_classical,schneider_1992_grav}.


By denoting the domain wall profile in spherical coordinate $\mb_0(\br) = (0, \sin\theta_0(x), \cos\theta_0(x))$ with $\theta_0$ the polar angle of magnetization $\mb_0$ with respect to $\hbz$, it is straightforward to find that the fictitious magnetic field $\bb= \tilde{D} \theta'_0(x) \mb_0$
always points opposite (because $\theta'_0 < 0$, see \Figure{fig:bend_sch}(a)) to the magnetization $\mb_0$, and its strength controlled by the magnetization gradient $\theta_0'(x)$.
Apparently, the projection of the magnetic field on the $x$-$z$ ($y$-$z$) plane is (anti-)symmetric about the domain wall center, \ie $b_y$ is negative in the whole region but $b_z$ is positive/negetive in left/right region, as depicted in \Figure{fig:bend_sch}(a).
In the meantime, the scalar potential $\phi$ inside the domain wall is a potential well, as illustrated by the gray-scale background centered at the domain wall in Fig. \ref{fig:bend_sch}(a),
which gives rise to an electric field $e_x$ that is antisymmetric about domain wall.

With above knowledge that the domain wall manifests itself as fictitious fields $\be$ and $\bb$, we may treat the spin wave scattering governed by \Eq{eqn:eom_wp} as a negatively charged particle deflected by these fields.
The electrostatic (Lorentz) force flips as electric (magnetic) field reverses, thus the spin wave deflection pattern depends on the symmetry of these fictitious electromagnetic fields.
Nevertheless, once the spin wave packet moves away from the domain wall, it takes straight trajectories, therefore simple geometric relations between incident and out-going spin wave beam are expected.

For theoretical simplicity as well as experimental relevance, here we focus on two scenarios: i) the \emph{normally magnetized} thin film case with the easy-axis anisotropy perpendicular to the film, corresponding to the film plane being $x$-$y$ plane in \Figure{fig:bend_sch}(a); ii) the \emph{tangentially magnetized} thin film case with the easy-axis anisotropy lying in the film, corresponding to the film plane being the $y$-$z$ plane in Fig. \ref{fig:bend_sch}(a).
More specifically, Fig. \ref{fig:bend_sch}(b)(c) show the slice cut for these two scenarios.
The perpendicular (to the film plane) magnetic field in normally/tangentially magnetized films are $b_z$ and $-b_y$ respectively, which are anti-symmetric and symmetric, while the electric field $e_x$ for both cases are antisymmetric.

For the special case of spin wave incidenting normally on the domain wall (Fig. \ref{fig:bend_sch}(b, c)), the spin wave experiences two qualitatively distinct fates across domain wall:
in the normally magnetized film, the spin wave is shifted laterally (Fig. \ref{fig:bend_sch}(b)), due to the opposite Lorentz forces in the left/right domain wall region caused by the antisymmetric magnetic field;
while in the tangentially magnetized film, the spin wave is bent upward (Fig. \ref{fig:bend_sch}(c)), because of the symmetric magnetic field.

\subsection{Numerical results}

To analyze the spin wave scattering problem more systematically, we turn to the numerical calculations.
Here two types of numerical calculations are performed in parallel: the full scale micromagnetic simulation (see Appendix \ref{sec:mag_sim}) based on the original LLG equation \eqref{eqn:LLG} and the trajectory simulation based on the semiclassical equation \eqref{eqn:eom_wp}.

We assume that the domain wall takes the Walker profile with $\mb_0=(0 ,\sech(x/W), \tanh(x/W))$ or $\theta_0(x) = 2\arctan[\exp(-x/W)]$, where $W=\sqrt{A/K}$ is the characteristic domain wall width\cite{lan_spin-wave_2015,yu_magnetic_2016}.
The effect of DMI is only to pin the domain wall as a Bloch-type, and does not alter the profile.
Upon this magnetization profile, the scalar potential is $\phi(x)= 2K \sech^2(x/W)$, which is a potential well since the magnetic energy density $u_0$ is larger inside domain wall.
The field components that can influence the spin wave trajectories are the magnetic (electric) field lying in the out-of-plane (in-plane) direction of the film plane, which are calculated for the normally magnetized and tangentially magnetized cases as in the following:
\begin{subequations}
\label{eqn:em_ipop}
\begin{align}
 e_\ssf{NM} = e_\ssf{TM} &= e_x = -\frac{4 K}{W}\sech^2\frac{x}{W}\tanh\frac{x}{W},\\
 b_\ssf{NM} &= b_z =-\frac{D}{2AW} \sech\frac{x}{W}\tanh\frac{x}{W}, \\
 b_\ssf{TM} &= - b_y =\frac{D}{2AW} \sech^2\frac{x}{W},
\end{align}
\end{subequations}
where the correspondence between $b_\ssf{NM/TM}$ and $b_{y/z}$ follow the coordinate setting in Fig. \ref{fig:bend_sch}.

\begin{figure*}[ht]
 \centering
 \includegraphics[width=0.99\textwidth]{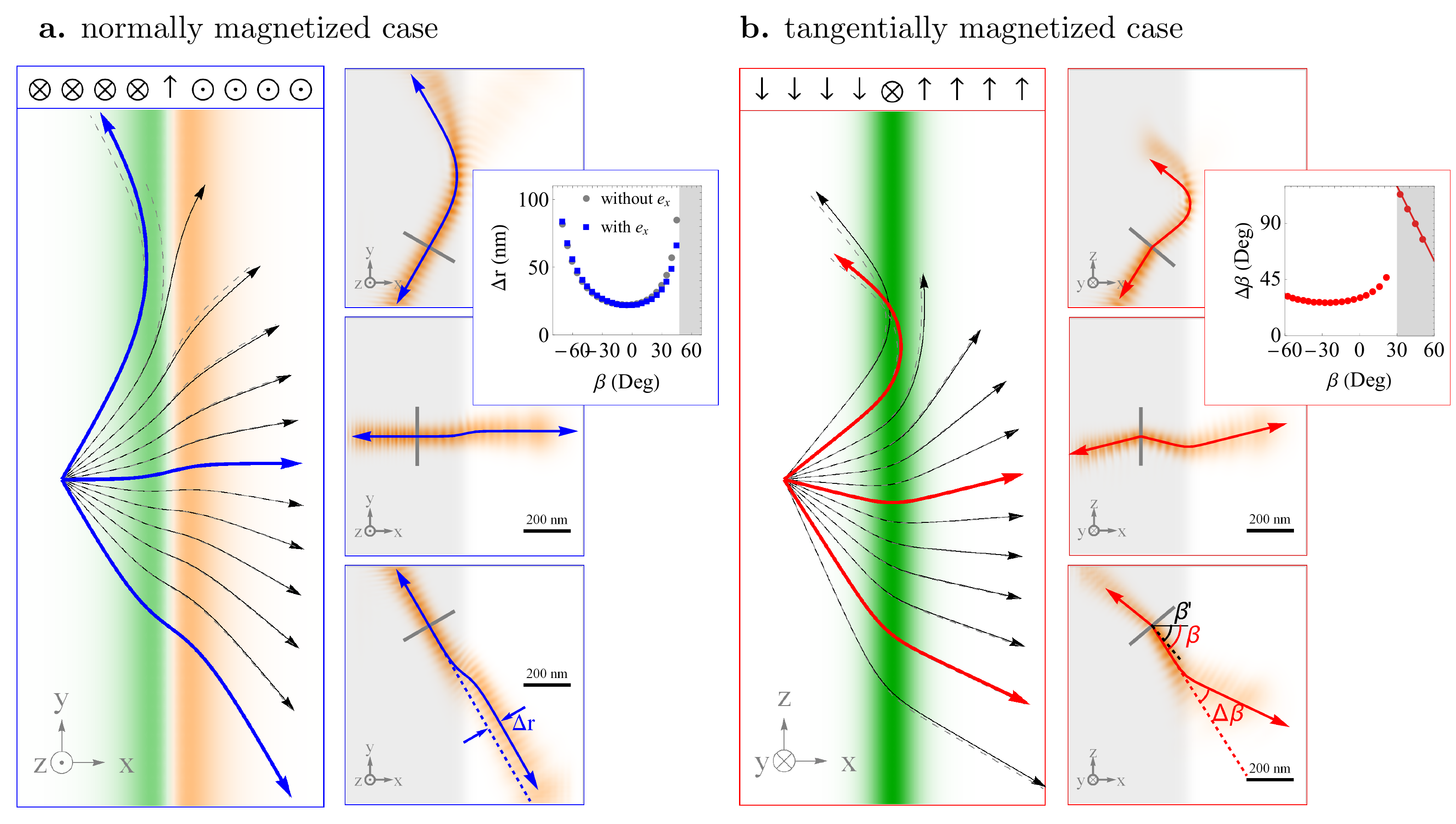}
 \caption{{\bf Numerical simulations of spin wave scattering by a chiral domain wall in (a) normally magnetized film and (b) tangentially magnetized film.}
In left panel, each line depicts a spin wave trajectory calculated from semiclassical equation \eqref{eqn:eom_wp} starting from the the same source point but with a specific incident angle, with solid/dashed lines denoting trajectories including/excluding the electric field.
The green/orange colors encode the positive/negative magnetic fields, and the arrows in the upper region denote the domain wall magnetizations.
In right panels, $3$ typical spin wave trajectories extracted from micromagnetic simulations are plotted in orange color, and the semiclassical trajectories are plotted in blue/red lines as in left panel.
The Gaussian spin wave beam is prepared in the gray antenna region with a discrepancy between beam direction and antenna direction in (b) (see Appendix \ref{sec:mag_sim}), and the spin wave trajectory are extracted based on spin wave flux (see Appendix \ref{sec:traj_track}).
Insets in (a)(b) plot the lateral shift $\Delta r$ and angle difference $\Delta \beta$ as function of incident angle $\beta$ respectively, and the gray area denotes the total reflection range.
In (b) inset, the gray/black dots are for the lateral shift with/without electric field; and in (b) inset, the solid line is the theoretical angle difference $\Delta\beta=180-2\beta$ for total reflection case.
For all numerical calculations and micromagnetic simulations, the spin wave frequency is $f=40~\mathrm{GHz}$, and the magnetic parameters are: exchange coupling constant $A=3.28\times 10^{-11} \mathrm{A}/\mathrm{m}$, anisotropy $K=3.88 \times 10^{5} \mathrm{A}/\mathrm{m}$, DMI constant $D=3\times 10^{-3}~\mathrm{A}$, and damping constant $\alpha=1\times 10^{-4}$.
}
 \label{fig:fm_sim}
\end{figure*}

With the fictitious electromagnetic fields $e_\ssf{NM/TM}$ and $b_\ssf{NM/TM}$ in \Eq{eqn:em_ipop}, the spin wave trajectories calculated from \Eq{eqn:eom_wp} with different incident angles are overlaid with the micromagnetic simulation results, as shown in Fig. \ref{fig:fm_sim} for the normally and tangentially magnetic film cases.
As expected, they agree well for all incident angles, and the out-going trajectory develops a lateral shift $\Delta r$ with respect to the incident trajectory in normally magnetized film, but forms an angle difference $\Delta\beta$ in tangentially magnetized film.
In addition, in both normally magnetized and tangentially magnetized films, when the incident angle exceeds a critical angle, the spin wave packet is totally reflected by the domain wall.
The reflection only occurs when spin wave incidents along the upward direction, i.e. $+\hby$ ($+\hbz$) in normally/tangentially magnetized film, highlighting the chiral nature of the underlying Lorentz force.
These trajectory shifting or bending behavior can be mostly understood from the magnetic field distributions.
However, the effective electric field also contributes in manipulating the spin wave trajectory, which is shown as the difference between the solid and dashed trajectories for including and excluding the effect of the electric field in the main panels of Fig. \ref{fig:fm_sim}(a)(b).
For most incident angles, the electrostatic force is dominated by the Lorentz force due to the large spin wave velocity, but its contribution becomes non-negligible around the total reflection situation.


The lateral shift $\Delta r$ in normally magnetized film and the angle different $\Delta \beta$ in tangentially magnetized film, as function of incident angle $\beta$, are summarized in inset of Fig. \ref{fig:fm_sim}(a, b) respectively.
Typically, as spin wave deviates from the normal incident direction of the domain wall, the velocity $v_x$ decreases and the passing time increases, thus both the lateral shift $\Delta r$ and angle different $\Delta\beta$ increase.
However, these two geometric quantities $\Delta r$ and $\Delta \beta$ are both asymmetric with respect to the incident angle $\beta$, since the spin wave is subject to chiral Lorentz force inside domain wall.
In Fig. \ref{fig:fm_sim}(a) inset, the lateral shift $\Delta r$ including/excluding electric field shows a discrepancy, highlighting the role of electrostatic force in developing lateral shift.
And in Fig. \ref{fig:fm_sim}(b) inset, the angle difference $\Delta\beta$ maximizes for a certain positive incident angle and start to decrease linearly, indicating the emergence of the total reflection of the spin wave beam.

\begin{figure}[t]
 \centering
 \includegraphics[width=0.48\textwidth]{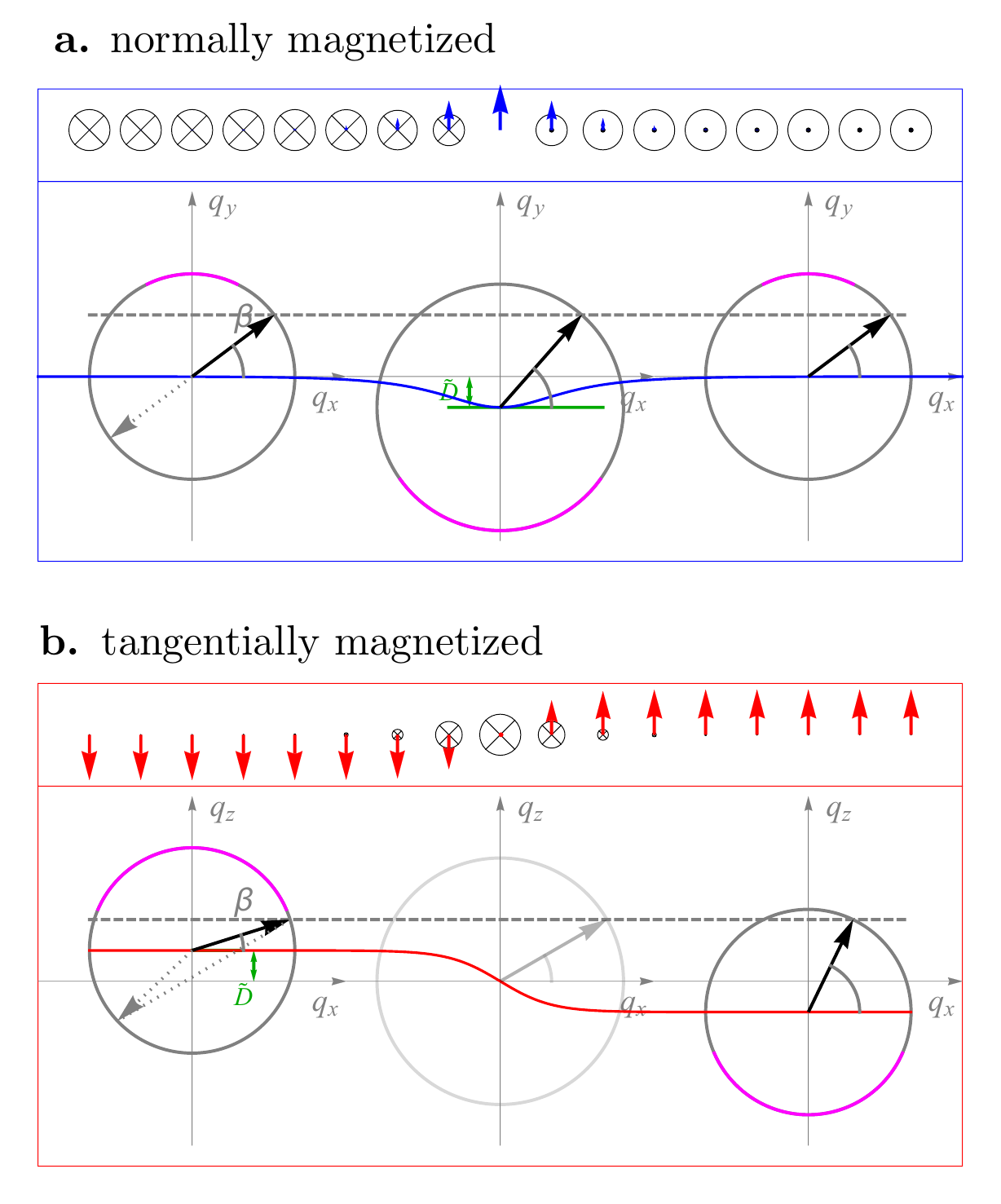}
 \caption{
 {\bf Schematics of magnonic Snell's law across chiral domain wall in (a) normally magnetized film and (b) tangentially magnetized film}.
 The isofrequency circles in the wavevector space $(q_x, q_{y/z})$ are plotted at the left/right domains and the domain wall center respectively.
 The blue/red line plots the profile of in-plane magnetization $m^{y/z}_0$, which acts as the generalized refraction index.
 The black arrow denotes the local momentum vector $\bk$, which forms angle $\beta$ with $x$ axis, and the evolution of angle $\beta$ are connected by dashed lines.
 The magenta arcs describe the modes with/without corresponding propagation modes in other regions.
 The dotted arrow represents the momentum $\bk$ of spin wave generated in the other side of the antenna.
 In upper region, the magnetization profile is depicted by arrows, with the in-plane component highlighted by blue/red colors.
 }
\label{fig:snell}
\end{figure}

\subsection{Magnonic Snell's law}
For a straight domain wall under consideration in this work, because of the translational invariance along $y/z$ axes, the wavevector $q_{y/z}$ is conserved.
Note that the canonical wavevector $\bk =\bq +\ba$, the angle $\beta$ formed between the spin wave beam and the normal direction of the domain wall obeys the following generalized magnonic Snell's law:
\begin{subequations}
\label{eqn:mag_snell}
\begin{align}
 \text{normally magnetized:}& \quad k\sin\beta - D m_0^y = \mathrm{const.},\\
 \text{tangentially magnetized:}& \quad k\sin\beta - D m_0^z = \mathrm{const.},
\end{align}
\end{subequations}
where the in-plane magnetization component ($m_0^{y/z}$ for normally/tangentially magnetized film) plays the role of generalized refraction index characterizing the magnetic medium.
The magnonic Snell's law in \Eq{eqn:mag_snell} holds everywhere inside the continuum medium, thus is an extension of previously proposed Snell's laws that only concerns two sides of an interface\cite{yu_magnetic_2016, stigloher_Snell_2016, mulkers_tunable_2018, hioki_snells_2020}.

The magnonic Snell's law formulated in \Eq{eqn:mag_snell} is schematically illustrated by the matching of corresponding iso-frequency circles, as depicted in Fig. \ref{fig:snell}.
Three representative positions are focused: the left/right domain with $\mb_0=\mp\hbz$ and the domain wall center $\mb_0=+\hby$.
For each isofrequency circle, the center is shifted to $\bq=-\ba = -\tilde{D} \mb_0$, and the radius is $k(x) = \sqrt{(\omega/\gamma)-K+ \phi(x)}$.
Specifically for the normally magnetized case in Fig. \ref{fig:snell}(a), the in-plane magnetization $m_0^y$ maximizes at the domain wall center and vanishes in left/right domains, therefore the domain wall mimics a three-layer system with low/high/low refraction indices.
Consequently, the spin wave experiences a lateral shift, similar to the lateral shift of light ray as passing through an air/glass/air structure.
As for the tangentially magnetized cases in Fig. \ref{fig:snell}(b), the in-plane magnetization $m_0^z$ monotonically decreases along $x$ direction, therefore the domain wall mimics a two-layer structure with low/high refraction indices, giving rise to the spin wave refraction,  similar to the case of light refraction in an air/water interface.
And since there is an interface of effectively low/high refraction indices for both normally and tangentially magnetized cases, the spin wave total reflection arises due to lacking of corresponding propagation mode in the other regions.

\begin{figure*}[t]
 \centering
 \includegraphics[width=\textwidth]{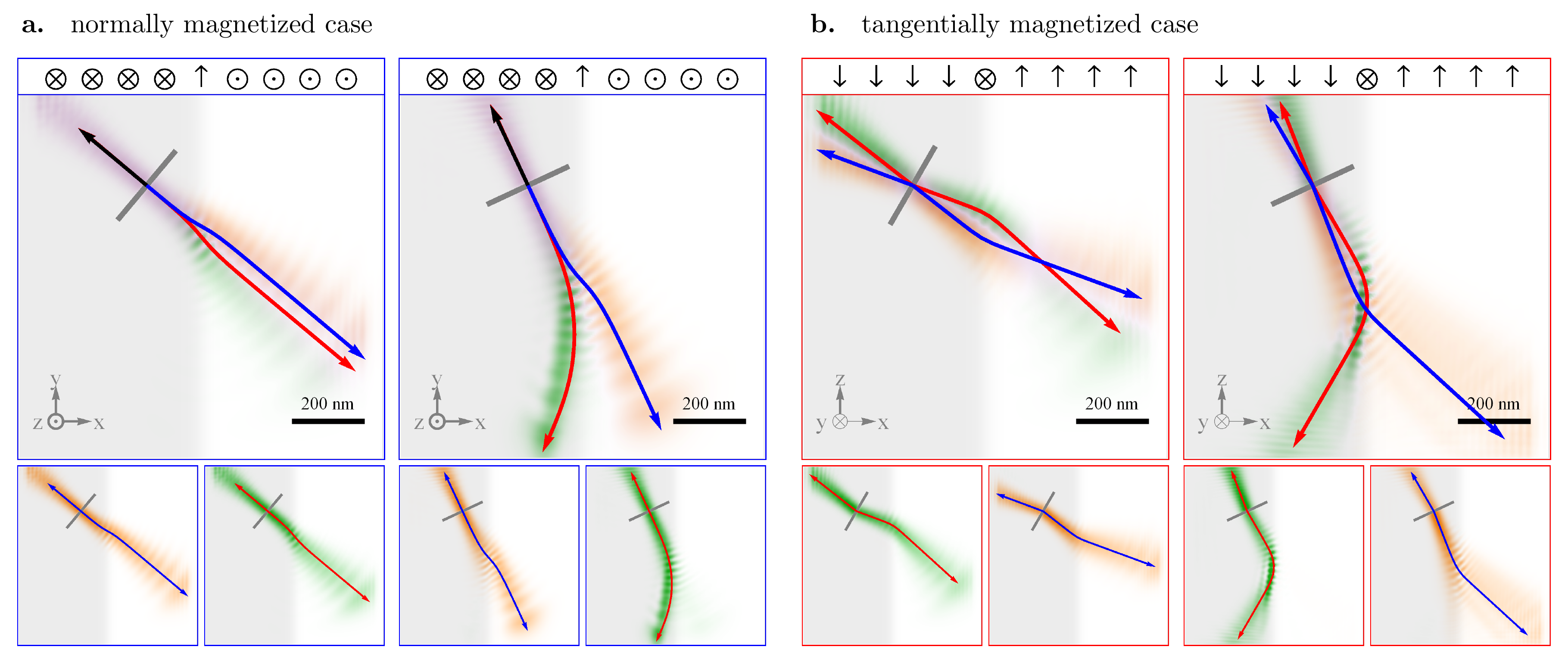}
\caption{
 {\bf Numerical simulations of spin wave scattering by an antiferromagnetic domain wall in (a) normally magnetized film and (b) tangentially magnetized film.}
In each main panel, the green/orange/purple color plots the trajectory of left/right circular and linear spin wave extracted from micromagnetic simulations, and the blue/red/black lines are corresponding trajectories calculated from semiclassical equation \eqref{eqn:wp_eom_AFM}.
The lower $2$ panels plots the trajectories of left/right circular spin wave separately.
A linearly polarized oscillating magnetic field is exerted in the antenna region (gray rectangle) at the domain wall center to generate spin wave.
The spin wave trajectory with polarization information are based on extraction of spin wave flux (see Appendix \ref{sec:traj_track}).
In all numerical simulations, the spin wave frequency is $f=50~\mathrm{GHz}$, and the magnetic parameters are: exchange coupling constant $A=3.28\times 10^{-11} \mathrm{A}/\mathrm{m}$, anisotropy $K=3.88 \times 10^{5}~ \mathrm{A}/\mathrm{m}$, DMI constant $D=2\times 10^{-3}~\mathrm{A}$, $J=1\times 10^6 ~ \mathrm{A}/\mathrm{m}$ and damping constant $\alpha=1\times 10^{-4}$.
}
\label{fig:afm_sim}
\end{figure*}

\section{Spin wave deflections by antiferromagnetic domain wall}

The spin wave deflections in ferromagnetic environment discussed above naturally extend to antiferromagnets, and their features are enriched by the additional polarization degree of freedom.
In antiferromagnets, due to two sublattices with opposite magnetizations, there exists both left/right circular polarization modes for spin wave \cite{cheng_antiferromagnetic_2016,lan_antiferromagnetic_2017-1,gitgeatpong_nonreciprocal_2017-2,nambu_observation_2020,li_spin_2020}.
Since these two circular modes precess in opposite fashion, they experience opposite fictitious magnetic fields induced by DMI, thus are deflected in opposite directions.

Here we denote the magnetization in two sublattices of antiferromagnets as $\mb_{1/2}$, then the staggered magnetization is $\bn=(\mb_1-\mb_2)/|\mb_1-\mb_2|$, and the net magnetization is $\mb=\mb_1+\mb_2$.
Under the approximation $\bn\cdot \mb=0$, the magnetic dynamics in antiferromagnets is governed by an LLG-like equation \cite{haldane_nonlinear_1983-1,tveten_staggered_2013-2,kim_propulsion_2014-3,tveten_intrinsic_2016,yu_polarization-selective_2018}
\begin{align}
 \label{eqn:LLG_AFM}
 \frac{1}{\gamma J} \bn \times \ddot{\bn} = - \gamma \bn \times \bh +\alpha \bn \times \dot{\bn} ,
\end{align}
where $\bh=A\nabla^2\bn+K n_z\hbz-D \nabla\times \bn$ is the effective field taking similar form as in \Eq{eqn:LLG}, and $J$ is the inter-sublattice exchange coupling constant.
And similarly, the total magnetization $\bn$ divides into the static background $\bn_0$ and the dynamical antiferromagnetic spin wave excitation $\delta \bn$: $\bn=\bn_0+\delta \bn$, with $\delta \bn=n_\theta \hbe_\theta+n_\phi \hbe_\phi$.
The domain wall has the same magnetization profile as in ferromagnetic case\cite{tveten_intrinsic_2016,yu_polarization-selective_2018}, and for spin wave dynamics, \Eq{eqn:LLG_AFM} is reduced to a Klein-Gordon-like equation \cite{yu_polarization-selective_2018,kim_tunable_2019}
\begin{align}
\label{eqn:sw_eom_AFM}
 -\ddot{\psi}_s = \gamma^2 J\qty[ (-i\nabla+s \ba)^2 +K- \phi]\psi_s,
\end{align}
where $\psi_s=n_\theta - i s n_\phi$ denotes the left/right circularly polarized spin wave with $s=\mp 1$ the chirality, and potentials $\phi$ and $\ba$ follow definitions in \Eq{eqn:sw_eom_FM}.
Following similar procedures as in \Eq{eqn:eom_wp}, the spin wave dynamics is recast from \Eq{eqn:sw_eom_AFM} to
\begin{equation}
\label{eqn:wp_eom_AFM}
 m_\ssf{AFM}\ddot{\br} = -\be - \dot{\br}\times s\bb,
\end{equation}
where the right/left circular spin waves take analogy to charged particles travelling in the same electric field $\be = -\nabla \phi$ and opposite magnetic fields $\mp \bb$ with $\bb= \nabla\times \ba$.  We should note that only the effective magnetic field changes sign for the two polarizations, the effective electric field is the same, therefore the two circular polarizations does not corresponds to the positive/negative charge.
Here the local spin wave dispersion is $\omega= \gamma\sqrt{ J(K+Ak^2)}$, the group velocity $\bv =\partial_\bk \omega= \gamma J A k/\omega$, and the effective mass $m_\ssf{AFM}=1/\partial_\bk^2 \omega=\omega^3/(\gamma J^2AK)$.


The polarization-dependent trajectories calculated from the semiclassical equation \Eq{eqn:wp_eom_AFM} and simulated based on micromagnetics are depicted in Fig. \ref{fig:afm_sim}, and they agree well with each other as expected.
In normally magnetized case (\Figure{fig:afm_sim}(a)), a linearly polarized spin wave beam is injected by the antenna.  For a large incident angle, the left/right circular modes beams experience opposite lateral shift and splits into two parallel beams, causing a double refraction. For small incident angles, one of the polarization would bend so much that it is totally reflected by the domain wall, while the other polarization still experience a lateral shift and penetrates into the other domain.
In tangentially magnetized case (\Figure{fig:afm_sim}(a)), the left/right circular polarizations of the same frequency do not have the same wavevector direction, thus they splits as soon as they leave the antenna. As they hit the domain wall, they are bending in opposite directions due to the opposite effective magnetic fields in the domain wall region, and total reflection can also happen for one of the polarizations if the incident angle is smaller than a critical angle.
All these polarization-dependent scattering patterns shown in Fig. \ref{fig:afm_sim} can be straightforwardly understood by the effective electromagnetic fields as in the ferromagnetic case in \Figure{fig:fm_sim}, or by extending magnonic Snell's law to antiferromagnetic environment, by using $\mp n_0^{y/z}$ as the as the generalized refraction indices for left/right circular modes.

The spin wave birefringence phenomenon observed in Fig. \ref{fig:afm_sim} refers to the polarization-dependent trajectories in 2D magnetic film, which is different from the polarization-dependent phase demonstrated in 1D magnetic wire in previous reports \cite{cheng_antiferromagnetic_2016, lan_antiferromagnetic_2017-1, han_Birefringencelike_2020}.
Recently, the bi-reflection of spin wave induced by the hybridization with elastic wave is also reported, where the film boundary rather than a domain wall serves as the scattering interface \cite{hioki_Bireflection_2020}.

\section{Spin wave constriction by domain wall}

\begin{figure}[t]
 \centering
 \includegraphics[width=\columnwidth]{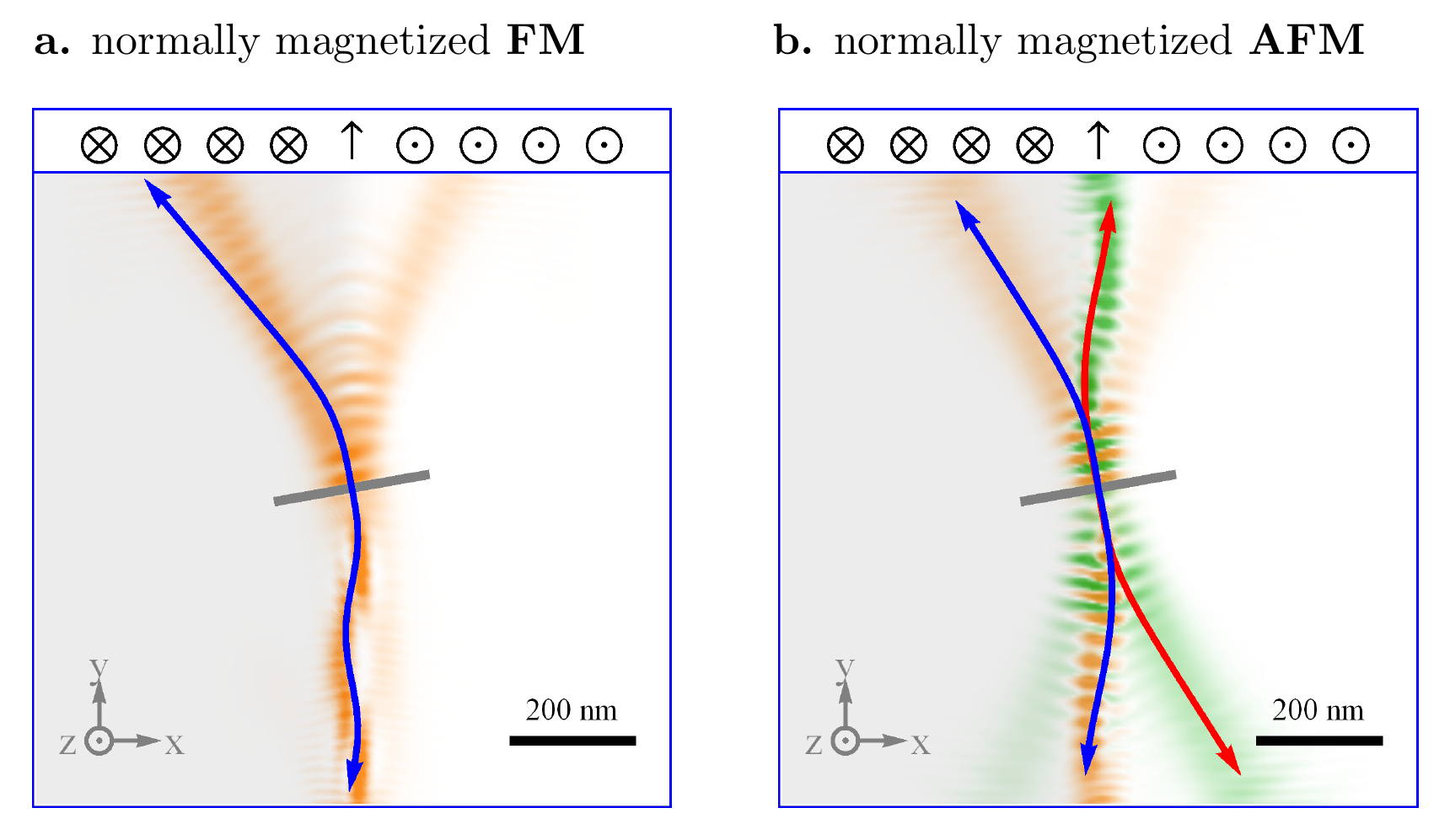}
\caption{{\bf Constriction of spin wave by a chiral domain wall in normally magnetized film in (a) ferromagnets and (b) antiferromagnets.}
The green/orange color plots the trajectory of left/right circular and linear spin wave extracted from micromagnetic simulations, and the blue/red lines are corresponding trajectories calculated from semiclassical equation \eqref{eqn:wp_eom_AFM}, and all other settings follows Fig. \ref{fig:afm_sim}.
The spin wave leaking in the upper-right in (a) is due to the sub-wave spreading in different angles for spin wave generated in the antenna, and similar leaking also occurs in (b).
}
\label{fig:sim_bound}
\end{figure}

 We have seen that the normally magnetized case can be considered as an analogy of air/glass/air for light (see Fig. \ref{fig:snell}), where the domain wall serves as the middle high refraction index ``glass'' layer. It is known that light can  be confined in the glass and travel along the glass layer without leaking into the air because of the total internal reflection, as widely used in optical fiber. In the normally magnetized film, a domain wall can also be used to guide spin waves just as glass guiding light.
\Figure{fig:sim_bound} shows exactly this phenomena in normally magnetized ferromagnetic and antiferromagnetic films.
In ferromagnetic case, when the spin wave is excited within the domain wall with a shallow incident angle, the downward-going spin wave is constricted within the domain wall with a snake-like trajectory, while the upward-going spin wave leaks into the bulk domains. Therefore, this spin wave constriction is unidirectional. This constriction is due to the opposite effect Lorentz force due to the opposite effective magnetic fields at the two sides of the domain wall (see the main figure in \Figure{fig:fm_sim}(a)). The AFM case is quite similar, but the constriction depends on the spin wave polarization.
This unidirectional constriction can be also understood using the isofrequency circle mismatching in Fig. \ref{fig:snell}(a).


This unidirectional constricted spin wave mode is different from the previously known spin wave bound state  \cite{garcia-sanchez_narrow_2015, lan_spin-wave_2015, wagner_magnetic_2016, sluka_emission_2019, wintz_magnetic_2016}.
There are two major differences: i) the spin wave bound state previously discussed has frequency lower than the bulk spin wave gap in the domains, while the constricted spin wave mode discussed above has frequency above the spin wave gap; ii) the spin wave bound state exists because of the scalar potential $\phi$ is a potential well (or the asymmetric effective electric field) and regardless of the vector potential or the effective magnetic field, while the constricted spin wave mode here exists only because of the Lorentz force caused by the asymmetric effective magnetic field.


In this work, the spin wave fiber is based on a single domain wall in normally magnetized film, thus is also different from the previously reported spin wave fiber relying on two parallel magnetic domain walls by present authors \cite{yu_magnetic_2016}.
Xing et al \cite{xing_fiber_2016} also reported the fiber-like spin wave propagation behavior within a chiral domain wall, but their results are based on single mode spin wave without demonstration of the total internal reflection.

%

\section{Conclusions}

In conclusion, we demonstrate that the spin wave scattering by a chiral domain wall can be simplified to geometric relations between incident and out-going beams.
Underlying these geometric scattering behavior is the deflection of spin wave by fictitious electromagnetic field induced by domain wall, where the deflection chirality is a collaboration of the chirality of DMI, domain wall and spin wave.
The geometric magnonics demonstrated in this work offers us new designing principles in controlling spin wave propagation and separating spin waves of opposite chiralities.

\acknowledgements{
This work is supported by  National Natural Science Foundation of China (No. 11904260), and the Startup Fund of Tianjin University.
W.C. is also supported by the JSPS Kakenhi (Grant No. 20K14369).
J.X. is also supported by the National Natural Science Foundation of China (No. 11722430, No. 11847202).
J.L. is grateful to Xinghui Feng for discussions about the eikonal equations.
}

\appendix
\section{Micromagnetic simulations}
\label{sec:mag_sim}
The micromagnetic simulations are performed in COMSOL Multiphysics, where the LLG equation is transformed into weak form and then solved by the generalized-alpha method \cite{COMSOL}.
In each simulation, a domain wall is placed at the magnetic film center, and a spin wave beam is prepared in the rectangle antenna region and incident on the domain wall.
Near the film boundaries, the damping constant $\alpha$ is gradually increased from $1\times 10^{-4}$ to $2\times 10^{-1}$ in $50\mathrm{nm}$ to absorb undesired spin wave.

To generate spin wave beam, the excitation magnetic field is set to take the Gaussian form \cite{gruszecki_influence_2015,yu_magnetic_2016,gruszecki_goos-hanchen_2017}
\begin{align}
h_\ssf{ex}=&h_0\cos(2\pi f t) \exp\qty(\frac{(x'-x_c')^2}{2\lambda^2})\nonumber \\
 \times& \Theta\qty(\frac{w_a}{2}-|x'-x_c'|)\Theta\qty(\frac{h_a}{2}-|y'-y_c'|),
\end{align}
where $\Theta(x)$ is the Heaviside step function, $h_0$ and $f$ are the strength and frequency of the excitation magnetic field, and $\lambda$ is the Gaussian distribution width.
Here $x'$ and $y'$ are the positions along the width/height direction of the antenna, $x'_c$ and $y'_c$ are the central positions of the antenna, $w_a$ and $h_a$ are the width/height of the antenna.
For micromagnetic simulations in this work, the antenna size is set to $w_a=250~\mathrm{nm}$, $h_a=15~\mathrm{nm}$, and the Gaussian distribution width is $\lambda=60~\mathrm{nm}$.

We denote the velocity angle $\beta$ as the angle between the propagation direction of spin wave beam and $x$-axis, and antenna angle $\beta'$ as the angle between the normal direction of antenna and $x$-axis. The velocity angle $\beta$ is then related to velocity $\bv$ (or canonical momentum $\bk$), and the antenna angle $\beta'$ is related to the wavevector $\bq$, with
\begin{align}
 \beta = \arccos\frac{v_x}{x}, \quad \beta'= \arccos \frac{q_x}{q}.
\end{align}
Due to the vector potential $\ba$ experienced by the spin wave packet, or the relation $\bk=\bq+\ba$, the velocity angle $\beta$ and antenna angle $\beta'$ are not necessarily the same.
More explicitly, the vector potential $\ba$ vanishes (maintains) in the uniformed domains in normally/tangentially magnetized films, thus the velocity angle $\beta$ equals to (deviates from) the antenna angle $\beta'$, i.e. $\beta=\beta'$ in normally magnetized case while $\beta\neq \beta'$ in tangentially magnetized case.

The antiferromagnetic simulations are performed in a synthetic antiferromagnetic film consisting of two ferromagnetic layers that are coupled antiferromagnetically \cite{yu_polarization-selective_2018}.
Denoting $\mb_{1/2}$ as the magnetization in upper/lower magnetic layer, then the magnetic dynamics is governed by coupled LLG equations
\begin{align}
\label{eqn:LLG_syAF}
 \dot{\mb}_i=-\gamma\mb_i\times \bh_i +\alpha \mb_i \times \dot{\mb}_i,
\end{align}
where $\bh_{i}= A \nabla^2 \mb_{i} + K m_{i}^z - J \mb_{\bar{i}}/2$ is the effective fields acting on $\mb_{i}$ with $\bar{1}=2, \bar{2}=1$ .
Defining staggered magnetization $\bn=(\mb_1-\mb_2)/|\mb_1-\mb_2|$, net magnetization $\mb=(\mb_1+\mb_2)$, and using the approximation $\bn\cdot \mb=0$, \Eq{eqn:LLG_syAF} is then recast to \Eq{eqn:LLG_AFM}.

\section{Trajectory tracking of spin wave beam}
\label{sec:traj_track}
To visualize the trajectory with polarization information of the spin wave beam, we define the local spin wave flux
\begin{align}
j(\br,t) = \mb_0(\br)\cdot (\dot{\mb}(\br,t)\times \mb(\br,t)),
\end{align}
where $\mb_0$ is the static magnetic background at $t=0$, and $\mb$ is the total magnetization at the time $t$ under consideration.
The spin wave flux $j$ is only nonzero when local magnetization precesses, and its sign is directly determined by chirality denoting the precession direction.
For right circular spin wave in ferromagnets, the corresponding flux $j$ is always negative, as shown in Fig. \ref{fig:fm_sim} and Fig. \ref{fig:sim_bound}(a).

In antiferromagnets, the polarized spin wave generally have both left/right circular polarization components, and their mixture complicated the trajectory analysis.
However, by observing that the left/right circular spin wave mainly resides at the upper/lower layer of the synthetic antiferromagnet, we may define layer-resolved spin wave flux
 \begin{align}
j_i(\br,t)= \mb^0_i(\br)\cdot (\dot{\mb}_i(\br,t)\times \mb_i(\br,t)),
\end{align}
where $i=1,2$ refers to the upper/lower layer.
With flux $j_{1/2}$ in upper/lower layer, the total flux $j=j_1+j_2$, and polarized flux $j'=(j_1-j_2)/2$ are used to depict the spin wave trajectory.
The signal of total flux $j$ maximizes for circular spin wave, and the signal of $j'$ maximizes for linear spin wave.

\end{document}